\documentstyle[aps,prl,preprint,psfig]{revtex}
\begin{document}

\title{Accurate first-principle  equation of state for 
the One-Component Plasma}

\author{Nikolai~V.~Brilliantov$^{1,2}$}
\address{$^{1}$Chemical Physics Theory Group, Department of Chemistry, 
University of Toronto, Toronto, Canada M5S 3H6}
\address{$^{2}$Moscow State University, Physics Department,
Moscow 119899, Russia}
\maketitle

\bigskip
\begin{abstract}

Accurate "first-principle" expressions  for the excess free 
energy $F_{\rm ex}$ and  
internal energy $U_{\rm ex}$ of 
the classical one-component plasma (OCP) are obtained. 
We use the Hubbard-Schofield transformation that maps the OCP Hamiltonian
onto the Ising-like Hamiltonian, with  coefficients  expressed in 
terms of equilibrium correlation functions of a  reference system. 
We use the ideal gas as a reference system for which all the 
correlation functions are known. Explicit calculations are performed with the 
high-order terms in the Ising-like Hamiltonian omitted. 
For small values of the plasma parameter $\Gamma$ the Debye-Huckel
result for $F_{\rm ex}$ and $U_{\rm ex}$ is recovered. For $\Gamma \gg 1$,
these depend linearly on $\Gamma$ in accordance with the Monte Carlo findings
for the OCP. The MC data for the internal energy are reproduced 
fairly well by the obtained analytical expression. 

\end{abstract}

\section{Introduction}
 The one component plasma (OCP) model is one of the basic models in the 
condensed matter physics \cite{1,2}. Besides of its  direct astrophysical 
applications \cite{2}, to model the  
ionized matter in white dwarfs \cite{3}, outer layers of 
neutron stars and  interiors of heavy planets \cite{4,5}, OCP is widely 
used as a reference model for a variety of systems, ranging from 
alkali metals \cite{6,7,8} to colloidal solutions \cite{9,10,11}. 
The application of the OCP is not restricted by classical systems; 
it is also used when quantum effects are important \cite{quan}.

The OCP
model is formulated as a system of point particles, interacting via the 
Coulombic potential, which move in a uniform neutralizing background 
\cite{1,2}. All thermodynamic properties of the OCP depend only on the   
dimensionless plasma parameter $\Gamma=l_B/a_c$, where 
$l_B=e^2/ k_BT$ is the Bjerrum length ($e$ is the charge of the 
particles, $k_B$ is the Boltzmann's constant, $T$ is the temperature) and 
$a_c=(3/4 \pi \rho)^{1/3}$ is the ion-sphere radius with 
$\rho=N/\Omega$ being the 
concentration of  particles ($N$ is the number of particles, 
$\Omega$ is the volume of the system). 

For small values of $\Gamma$, which correspond to a hot and/or dilute 
system, the Debye-Huckel theory accurately describes the 
thermodynamic properties of the OCP. 
For the excess free energy  and internal energy  this 
gives in the  $\Gamma \to 0$ limit:
%%%%%%%%%%%%%%%%%%%%%%%%
\begin{equation}
\label{DH}
\frac{F_{\rm ex}}{k_BTN}=-\frac{1}{\sqrt{3}}\Gamma^{3/2};\, \, \, \,\,\,
\frac{U_{\rm ex}}{k_BTN}=-\frac{\sqrt{3}}{2}\Gamma^{3/2}
\end{equation}
Abe expansion \cite{12} for the classical OCP provides next 
terms for that limit (in the present study we 
 consider the classical case):
 
\begin{eqnarray}
\label{abeF}
\frac{F_{\rm ex}}{k_BTN}&=&
- \frac{1}{\sqrt{3}} \Gamma^{3/2}-\frac{c}{3}\Gamma^{3}-
\frac{1}{8} \Gamma^{3} \left( 3 \log \Gamma -1\right)+\cdots \\ 
\label{abeU}
\frac{U_{\rm ex}}{k_BTN}&=&
- \frac{\sqrt{3}}{2} \Gamma^{3/2}-
\left( c+\frac{3}{8} \right) \Gamma^{3} - 
\frac{3}{8} \Gamma^{3}\left( 3 \log \Gamma -1\right)+ \cdots
\end{eqnarray}
where $c=\frac98 \log 3 +\frac32 \gamma -1=1.101762 \ldots$, and 
$\gamma$ is the Euler's constant.
The analytical Abe expansion seems to be fairly  accurate for 
$\Gamma $ up to $0.1$ \cite{15}. 
 The next few terms for  the $\Gamma \to 0$
expansion calculated in Ref.\cite{CM} allow to 
use the small-$\Gamma$ expansion  up to $\Gamma \le 0.4$.

For larger values of $\Gamma$ the OCP was studied 
numerically, by means of integral equations, such as  
Percus-Yevick,  hypernetted-chain equations \cite{spring}, and   (most 
successively) by modified hypernetted-chain equation \cite{19}. 
Extensive numerical studies have been performed by Monte Carlo (MC)
\cite{5,15,14a,14,16,17} and Molecular Dynamics \cite{18} technique. 
To fit available ``experimental'' data for  the excess thermodynamic 
functions simple analytical fits were proposed \cite{20}:

\begin{eqnarray}
\label{Ufit}
\frac{U_{\rm ex}}{k_BTN}&=&A \Gamma +B\Gamma^s +C \\
\label{Ffit}
\frac{F_{\rm ex}}{k_BTN}&=&A \Gamma +\frac{B}{s}\Gamma^s +(3+C)\log \Gamma-D
\end{eqnarray}
with  $A=-0.8992$, $B=0.596$, $s=0.3253$, $C=-0.268$ and 
$D=A+B/s+1.1516$. Eqs.(\ref{Ufit}) and (\ref{Ffit}) are fairly  accurate 
for the interval $1 \le \Gamma \le 220$ \cite{20}, but 
unfortunately they do not give a  correctly behavior at $\Gamma \to 0$. 
Pade approximants for the $U_{\rm ex}$ suggested  in 
\cite{brami,kahl} remedy 
the failure of (\ref{Ufit}) at small $\Gamma$; one thus obtains 
very precise description for the whole interval 
$0 \le \Gamma \le 200$ \cite{kahl}.

To describe the limit of very large $\Gamma$ a perturbation theory was 
proposed \cite{21}; it agrees well with the MC data for large $\Gamma$, 
but also lack the proper   small-$\Gamma$ behavior. 
The  correct Debye-Huckel behavior at  $\Gamma \to 0$ together with 
a reasonable $ \sim 10 \%$ accuracy for $0 \le \Gamma \le 100$ 
has been obtained in a simple 
semiphenomenological ``Debye-Huckel plus hole'' (DHH) theory \cite{22};
it fails, however, for $\Gamma >125$ \cite{22}. A modified DHH theory  
proposed recently \cite{werner} using only one fitting parameter 
accurately reproduces 
the MC results for the large values of $\Gamma$ 
($1<\Gamma <200$) and demonstrates a correct behavior at $\Gamma \to 0$.

Thus, up to now, no  ``first-principle'' theory of OCP  exists 
which describes accurately the thermodynamic properties in the 
whole range of 
$\Gamma$ from the Debye-Huckel limit $\Gamma \to 0$ up  to   
$\Gamma \gg 1$ limit, where the Wigner crystallization \cite{5} occurs. 
In the present study we report  a simple ``first-principle'' equation of 
state for the OCP which has  the correct  Debye-Huckel behavior for  
small $\Gamma$ and  demonstrates a linear dependence 
on $\Gamma$ for $\Gamma \gg 1$. It 
reproduces within $1-2\%$ accuracy the experimental data 
for the most of  range of $\Gamma$  ($0 \le \Gamma \le 250$) and has  
a typical deviation  of the order of $2-5\%$ 
(with a maximal one  $\approx 8 \%$) for $0.01 < \Gamma < 1$. 
The rest of the paper is organized as follows:~in the next Sec.II 
we consider the Hubbard-Schofield transformation that maps the OCP Hamiltonian
onto the Ising-like Hamiltonian and calculate the coefficients of the 
effective Hamiltonian. In this section we also present the field-theoretical 
formulation for the statistical sum of the OCP which directly follows 
from the transformation used. Within the Gaussian approximation for the 
effective Hamiltonian we derive the equation of state for the OCP. 
In Sec.III we discuss the equation of state obtained and compare the 
analytical results for the internal excess free energy with the 
available Monte Carlo data for the OCP. In the last Sec.IV we summarize our 
findings.

%%%%%%%%%%%%%%%%%%%%%%%%%%%%%%%%%%%%%%%%%%%%%%%%%%%%%%%%%%%%%%%%

\section{Effective Hamiltonian and Equation of State for the OCP}

We start from the OCP Hamiltonian which may be written 
as follows ($\beta^{-1}=k_BT$): 

\begin{equation}
\label{hammic}
H = 
\frac{1}{2} \beta^{-1}{\sum_{{\bf k}}}^{\, \prime} \nu_k
\left(\rho_{{\bf k}} \rho_{-{\bf k}}-\rho \right)+
H_{\rm id}  
\end{equation}
where the first term in the right-hand side of Eq.(\ref{hammic}) refers to 
the Coulombic interactions, written in terms of the collective variables,
\begin{equation}
\rho_{{\bf k}} = \frac{1}{\sqrt{\Omega}} \sum_{j=1}^N e^{-i{\bf k}{\bf r}_j}
 \nonumber 
\end{equation}
where ${\bf r}_j$ denotes coordinate of $j$-th particle, 
 $\nu_k=4\pi l_B /k^2$ and  
$H_{id}$ is  the ideal-gas part of the Hamiltonian. 
 Summation in Eq.(\ref{hammic}) is to be performed 
over the wave-vectors ${\bf k}= \{k_x, k_y, k_z \}$ with  
$k_{i}=\frac{2\pi}{L}l_i$ ($i=x,y,z$), where $l_i$ are integers,  
$L^3=\Omega$, and the prime over 
the sum denotes that the term with ${\bf k}=0$
is excluded~\cite{remarkid} .  

\subsection{Hubbard-Schofield transformation}

The configurational integral may be then written 
in terms of the configurational integral of the reference 
(ideal gas) system $Q_R$~\cite{hubbard,brill1} as 

\begin{equation}
Q=\left\langle \exp \left\{-\frac{1}{2}\, {\sum_{{\bf k}}}^{\,\prime} \nu_{k} 
\left( \rho_{{\bf k}} \rho_{-{\bf k}}-\rho \right) \right\} \right\rangle_R Q_R
\label{Q}
\end{equation}
where  
$\langle\left(\cdots \right)\rangle_R =Q_R^{-1} \int d{\bf r}^N\left(\cdots \right)$ denotes the averaging over the reference system.
In accordance with the Hubbard-Schofield scheme ~\cite{hubbard} we use
the identity,

$$
\exp(-\frac{1}{2} a^2x^2)=(2 \pi a^2)^{-1/2}
\int_{-\infty}^{+\infty} \exp (-\frac{1}{2} y^2/a^2 + ixy)dy
$$
and arrive after some algebra at:

\begin{equation}
\label{QR1}
Q = Q_R \int {\prod_{{\bf k}}}^{\prime} c_{{\bf k}} d\varphi_{{\bf k}}
\exp \left\{-\frac12 {\sum_{{\bf k}}}^{\,\prime} \nu_k^{-1}\varphi_{{\bf k}} 
\varphi_{-{\bf k}} \right \} 
\left \langle \exp \left\{ i \, {\sum_{{\bf k}}}^{\,\prime} \rho_{{\bf k}}
\varphi_{-{\bf k}} \right\} \right \rangle_R
\end{equation}
where $c_{{\bf k}}=\left( 2\pi \nu_k \right)^{-1/2} e^{\frac12 \nu_k \rho}$,
and where the integration is to be performed under the restriction,
$\varphi_{-{\bf k}}=\varphi_{{\bf k}}^*$ ($\varphi_{{\bf k}}^*$ is the complex
conjugate of $\varphi_{{\bf k}}$) \cite{remsico}.
Applying the cumulant  theorem ~\cite{cum} to the factor
$\left \langle \exp \left\{ i\,{\sum_{{\bf k}}}^{\prime} \rho_{{\bf k}}
\varphi_{-{\bf k}} \right\} \right \rangle_R$ 
one obtains:

\begin{eqnarray}
&&Q = Q_R \int {\prod_{{\bf k}}}^{\prime} c_{{\bf k}}
d\varphi_{{\bf k}}e^{-{\cal  H}}, 
 \mbox{~~~with~~} \nonumber \\
&&{\cal H} =\sum_{n=2}^{\infty} \Omega^{1-\frac{n}{2}}
{ \sum_{ {\bf k}_1, \ldots {\bf k}_n } }^{ \prime} 
u_n\left( {\bf k}_1, \ldots {\bf k}_n \right)
\varphi_{{\bf k}_1}\cdots\varphi_{{\bf k}_n}\nonumber \\ 
&&u_{2}\left( {\bf k}_1, {\bf k}_2  \right)=
\frac12 \, \delta_{ {\bf k}_1+{\bf k}_2, {\bf 0} }
\left\{ \frac{k_1^2}{4 \pi l_B} +
\left\langle \rho_{{\bf k}_1} \rho_{-{\bf k}_1} \right\rangle_{cR} \right\} 
\nonumber \\
&&u_n \left( {\bf k}_1, \ldots {\bf k}_n  \right)=
-i^n\,\frac{ \Omega^{\frac{n}{2}-1} }{n!} \left\langle \rho_{{\bf k}_1} 
\ldots \rho_{{\bf k}_n} \right\rangle_{cR} ~~~n>2
\label{effham}
\end{eqnarray}
here $\left\langle \ldots \right\rangle_{cR}$ denotes
{\it cumulant average} ~\cite{cum} for 
the reference ideal-gas system. As it follows from Eqs.(\ref{effham}),
$Q$ is written in a similar  way as the partition function for the
magnetic system having the Ising-like Hamiltonian \cite{goldenfeldma}, where
$\varphi_{ \vec{k} }$ are the Fourier components of the
``spin--field'', $\varphi( \vec{r} )$.
Note that the coefficients of the effective
Hamiltonian (\ref{effham}) are expressed in terms of the correlation
functions of the ideal-gas system, which are perfectly 
known (e.g. \cite{golovko}):

\begin{equation}
\label{cumul}
\Omega^{ \frac{n}{2}-1} \left\langle \rho_{{\bf k}_1} 
\ldots \rho_{{\bf k}_n}\right\rangle_{cR}
=\rho \, \delta_{{\bf k}_1+ \cdots {\bf k}_n, {\bf 0}}
\end{equation}
This yields the effective Hamiltonian:

\begin{eqnarray}
&&{\cal H} =\frac12 \,{\sum_{{\bf k}}}^{\, \prime}
\left[\rho + (4\pi l_B)^{-1} k^2 \right]\varphi_{{\bf k}} \varphi_{-{\bf k}} 
-\sum_{n=3}^{\infty} \frac{i^n \Omega^{1-\frac{n}{2}}}{n!}
\, \rho {\sum}^{\, \prime}_{{\bf k}_1, \ldots {\bf k}_n }
\varphi_{{\bf k}_1}\cdots\varphi_{{\bf k}_n}
\delta_{{\bf k}_1+ \cdots {\bf k}_n, {\bf 0}} \nonumber \\  
&&={\cal H}_{G}+{\cal H}_1
\label{effham1}
\end{eqnarray}
where we write explicitly the   Gaussian part ${\cal H}_{G}$:

\begin{equation}
\label{gauss}
{\cal H}_{G}=\frac12 \,{\sum_{{\bf k}}}^{\, \prime}
\left[\rho + (4\pi l_B)^{-1} k^2 \right]\varphi_{{\bf k}} \varphi_{-{\bf k}}
\end{equation}
Since all the coefficients of the effective Hamiltonian are known, one 
can develop the  usual perturbation expansion, with ${\cal H}_{G}$ 
being  a reference part of the Hamiltonian and with ${\cal H}_1$ 
being perturbation
(e.g.~\cite{goldenfeldma,khol,ort}). 

A simple structure of the effective Hamiltonian ${\cal H}$, 
Eq.(\ref{effham1}) suggests  
a simple, closed-form field-theoretical formulation for  the partition sum 
of the OCP, akin the Sine-Gordon representation of the Coulombic 
gas \cite{klein}, or the restricted primitive model of electrolytes
\cite{khol,brilBB}. 

\subsection{Field-theoretical model for the OCP}

 Using the space-dependent field $\varphi( {\bf r} )$
$$
\varphi( {\bf r})= \frac{1}{\sqrt{\Omega}} \sum_{{\bf k}} \varphi_{{\bf k}}
e^{-i{\bf k}{\bf r}}
$$
It is easy to show that under this transformation the terms 
in Eq.(\ref{effham1}) containing products 
$ \varphi_{{\bf k}_1}\cdots\varphi_{{\bf k}_n}$ give rise to the 
terms $ \int d {\bf r} \varphi^n( {\bf r} )$, the terms  
$k^2 \varphi_{{\bf k}}\varphi_{-{\bf k}}$ give  
rise to the term $ \int d {\bf r} \left( \nabla \varphi \right)^2$, 
so that one can write

\begin{equation}
{\cal H} = \rho \int d {\bf r}\left[  \frac12(4\pi l_B \rho)^{-1}
\left( \nabla \varphi \right)^2 - \sum_{n=2}^{\infty}
\frac{i^n}{n!} \varphi^n \right]
\end{equation}
and recognize the expansion of $e^{i\varphi}$. 

Integration over the Fourier-components 
$\varphi_{{\bf k}}$ in Eq.(\ref{effham}) 
converts into "field"-integration over the 
field $\varphi( {\bf r} )$. It may be shown that the Jacobian of this 
transformation  does not depend on $\varphi( {\bf r} )$ and appears 
as a normalization  constant. 

Noticing that 
$$
{\prod_{{\bf k}}}^{\,\prime}\left( 2\pi \nu_k \right)^{-\frac12}= 
\left[ \int { \prod_{{\bf k}} }^{\,\prime} d \varphi_{{\bf k}} 
\exp \left\{ -\frac12 \, {\sum}^{\,\prime}_{{\bf k}} \frac{k^2}{4 \pi l_B} 
\varphi_{{\bf k}} \varphi_{-{\bf k}} \right\} \right]^{-1}
$$
(with the restriction $\varphi_{{\bf k}}^*=\varphi_{-{\bf k}}$)
and that 
$$
{\prod_{{\bf k}}}^{\,\prime} e^{\frac12 \nu_k \rho}=
\exp \left\{\rho \Omega \,\,\frac{U(0)}{2k_BT} \right\}
$$
where ${U(0)}$ is the so-called "self-energy" \cite{klein,remarkdiv}, 
one arrives after some algebra at the field-theoretical expression of 
the statistical sum $Z_{OCP}$ of the OCP:

\begin{equation}
\label{SGl}
e^{-\beta F} \equiv Z_{OCP}=
\frac{\int {\cal D} \varphi \, \exp \left\{-{\cal H}(\varphi)\right\} }{\int {\cal D}\varphi \,
\exp \left\{-\int d {\bf r} \left[ \frac12 \kappa_D^{-2}(\nabla \varphi)^2-\beta \tilde{\mu} \right]\right\}}
\end{equation}
where ${\cal D} \varphi$ denotes the "field"-integration, 

\begin{equation}
{\cal H} = \int d {\bf r}\left[\frac{\kappa_D^{-2}}{2}
\left( \nabla \varphi \right)^2 + i \varphi -e^{i\varphi} \right],
\end{equation}
$\kappa_D^2=4\pi l_B \rho= 4\pi e^2 \beta \rho$ is   
the inverse Debye screening length and  

\begin{equation}
\tilde{\mu} =\mu_{\rm id}+k_BT-U(0)/2. 
\end{equation}
Here $\beta \mu_{\rm id}= \log \left(\Lambda^3/\Omega \right)$ is the 
ideal-gas chemical potential ($\Lambda=h/(2\pi m k_BT)^{1/2}$ 
is the thermal wavelength). Deriving Eq.(\ref{SGl}) we write the 
statistical sum of the ideal gas as 
$Z_{\rm id}=\exp \left[-N \log ( \Lambda^3/\Omega ) \right]= 
\exp \left\{-\beta \mu_{\rm id} \, \rho \int d{\bf r} \right\}$ and 
rescale the length $\rho d {\bf r} \to d {\bf r}$.  

As it follows from  Eq.(\ref{SGl}) the potential function of the 
effective field-theoretical Hamiltonian for the OCP reads 
$V(\varphi)=i\varphi -e^{i\varphi}$. This may be compared with the 
potential function $V_{SG}(\varphi)=\cos \varphi $ of the Sine-Gordon 
model for the Coulombic gas \cite{klein}. Note, that all imaginary 
terms in  Eq.(\ref{SGl}) vanish after the field integration due to the 
symmetry properties of the Hamiltonian under the transformation   
$ \varphi \to -\varphi$. 

Consider now the "saddle-point" approximation to the numerator in 
Eq.(\ref{SGl}). The equation for the "extremal" field which  minimizes 
the effective Hamiltonian reads:

\begin{equation}
\label{pb1}
\nabla^2 \varphi=\kappa_D^2 \left(i-ie^{i\varphi} \right).
\end{equation}
Under  transformation $\varphi =i e^2 \phi /k_BT$  Eq.(\ref{pb1})  converts into
\begin{equation}
\nabla^2 \phi=-4 \pi \rho \left[e^{-e \phi/k_BT} -1 \right],
\end{equation}
which is the usual {\it mean-field} Poisson-Boltzmann equation for the OCP. 
This is not surprising since the "saddle-point" approximation is essentially
the mean-field one.   

\subsection{Equation of state for the OCP }

Now  we concentrate on the Gaussian part of the effective Hamiltonian 
and show that even neglecting 
the non-Gaussian contribution to the effective Hamiltonian, one can 
obtain fairly accurate equation of state for the OCP, provided that 
a correct value of the ultraviolet cutoff in the ${\bf k}$-space is 
used. The Gaussian approximation to ${\cal H}$ corresponds actually to the 
Random Phase, or Debye-H\"uckel approximation, 
(e.g.~\cite{krall}). Using ${\cal H}_{G}$, Eq.(\ref{gauss}) and 
performing (Gaussian) integration (e.g.~\cite{klein}) 
 over $\varphi_{{\bf k}}$ in (\ref{effham}), one 
easily finds for the excess free energy of the OCP:

\begin{equation}
\label{fk}
-\beta F_{\rm ex}= \log (Q/Q_R)=\frac12 \, {\sum_{{\bf k}}}^{\prime}
\left[ \rho \nu_k -\log \left(1+ \rho \nu_k \right) \right]
\end{equation}
We argue that the summation in Eq.(\ref{fk}) should be carried
out over a {\em finite} number of the wave-vectors ${\bf k}$. 
In this we follow the 
Debye theory of the specific heat of solids (e.g. \cite{ziman}). 
Namely,  we assume that 
the total number of degrees of freedom in the system, $3N$, should be equal to the total 
number of {\em physically different} modes with the wave-vectors ${\bf k}$ 
within the spherical shell of radius $k_0$ in the ${\bf k}$-space. 
 The number of modes 
is twice the number of the wave-vectors, since for each ${\bf k}$ one 
has a sine and cosine mode (the amplitude of the ${\bf k}$-th mode 
is a complex number) \cite{remark1}. Thus we obtain:

\begin{equation}
\label{k0}
2 \, \frac{\Omega}{8 \pi^3} 4\pi \int_0^{k_0}k^2 dk =3\,N 
\end{equation}
where the factor $\Omega/8 \pi^3$ appears when the 
integration in ${\bf k}$-space 
is used instead of summation.  From Eq.(\ref{k0}) follows that 
$k_0=\left(9 \rho \pi^2 \right)^{1/3}$. A similar Debye-like scheme 
to find the cutoff $k_0$ was first proposed for plasma in \cite{kaklugin},
 where 
somewhat different value of the cutoff wave-vector was reported. 
Using the $k_0$ obtained we 
write:

\begin{eqnarray}
-\frac{\beta F_{\rm ex}}{N}&=&
\frac12 \frac{\Omega}{8 \pi^3} \frac{4 \pi}{N}\int_0^{k_0} k^2 dk 
\left[ \log \left(1+\frac{\kappa_D^2}{k^2} \right) -\frac{\kappa_D^2}{k^2}\right] \nonumber \\
&=&\frac94 \int_0^1x^2dx \left[ \log \left(1+\frac{b\Gamma}{x^2} \right) -\frac{b\Gamma}{x^2} \right]
\end{eqnarray}
where $b=\frac{2}{3}\left( \frac{2}{\pi^2} \right)^{1/3}$. 
The last integral is easily calculated to obtain for the 
 free energy 
\begin{equation}
\label{result}
\frac{F_{\rm ex}}{k_BTN}=\frac{3}{4}\left[ \log \left( 1 +b\Gamma \right)-
b\Gamma \right]-
\frac{3}{2}\left( b\Gamma \right)^{\frac{3}{2}} \arctan \left( 
\frac{1}{\sqrt{b \Gamma}} \right) \nonumber
\label{result}
\end{equation}
and for the internal energy: 
\begin{equation}
\label{resU} 
\frac{U_{ex}}{k_BTN}=
-\frac{9}{4}\left(b\Gamma \right)^{\frac{3}{2}} \arctan \left(
\frac{1}{\sqrt{b \Gamma}} \right)
\end{equation}
of the OCP. To obtain Eq.(\ref{resU}) we use the relation 
$U_{\rm ex}=\Gamma \, \partial F_{\rm ex} / \partial \Gamma$. Again we note that 
the same functional dependence for the excess internal energy (but with different 
coefficients) has been obtained  in \cite{kaklugin}.

\section{Results and Discussion}

As it follows from Eqs.(\ref{result}) and (\ref{resU}), for $\Gamma \to 0$ the 
Debye-Huckel behavior is recovered. On the other hand in the opposite limit 
$\Gamma \gg 1$ Eqs.(\ref{result}) and (\ref{resU}) demonstrate the linear 
behavior 
on $\Gamma$ with the leading term $-A\Gamma$ in accordance with fits 
(\ref{Ufit}) and (\ref{Ffit}) to the MC data. The constant $A$ reads

\begin{equation}
\label{A} 
A=\frac94 b = \frac32 \left( \frac{2}{\pi^2} \right)^{1/3}=0.881\ldots
\end{equation}
which is fairly close to the constant $A=0.899\ldots$ 
of the fits   (\ref{Ufit}),  (\ref{Ffit}).

In Fig.1 the excess internal energy given by Eq.(\ref{resU}) is compared 
to the 
Monte Carlo data, taken from Ref.\cite{15,14a} for 
$ 0.1 \le \Gamma \le 1$ and Ref.\cite{6,16} for $\Gamma >1$.
Fig.2 shows the relative error of the analytical expression 
(\ref{resU}). 
As it follows from Fig.1 and Fig.2 the equation of state is fairly 
accurate in the most of range of the plasma parameter. The maximal deviation of the 
analytical expression from the numerical data  occurs at  the  
intermediate values of the plasma parameter, $0.01< \Gamma <1$. 

To analyze the reason  of the enhanced deviation of 
the theoretical results from the 
numerical data  at $0.1 < \Gamma < 0.5$,  one can address the 
small-$\Gamma$ expansion 
of  $U_{\rm ex}$ \cite{werner}. It was observed \cite{werner}
 that in spite of the correct Debye-Huckel limit,  
this does not reproduce correctly the next-order terms of the Abe 
expansion (\ref{abeU}).  This occurs  due to limitations 
of the Gaussian approximation for the effective 
Hamiltonian. Since all the coefficients of the effective Hamiltonian 
are known, one can go beyond the Gaussian approximation and 
develop a usual perturbation scheme, based on the Gaussian
Hamiltonian. 
\vspace{1cm}

\begin{minipage}{16cm}
\begin{figure}[htbp]
  \centerline{ 
  \psfig{file=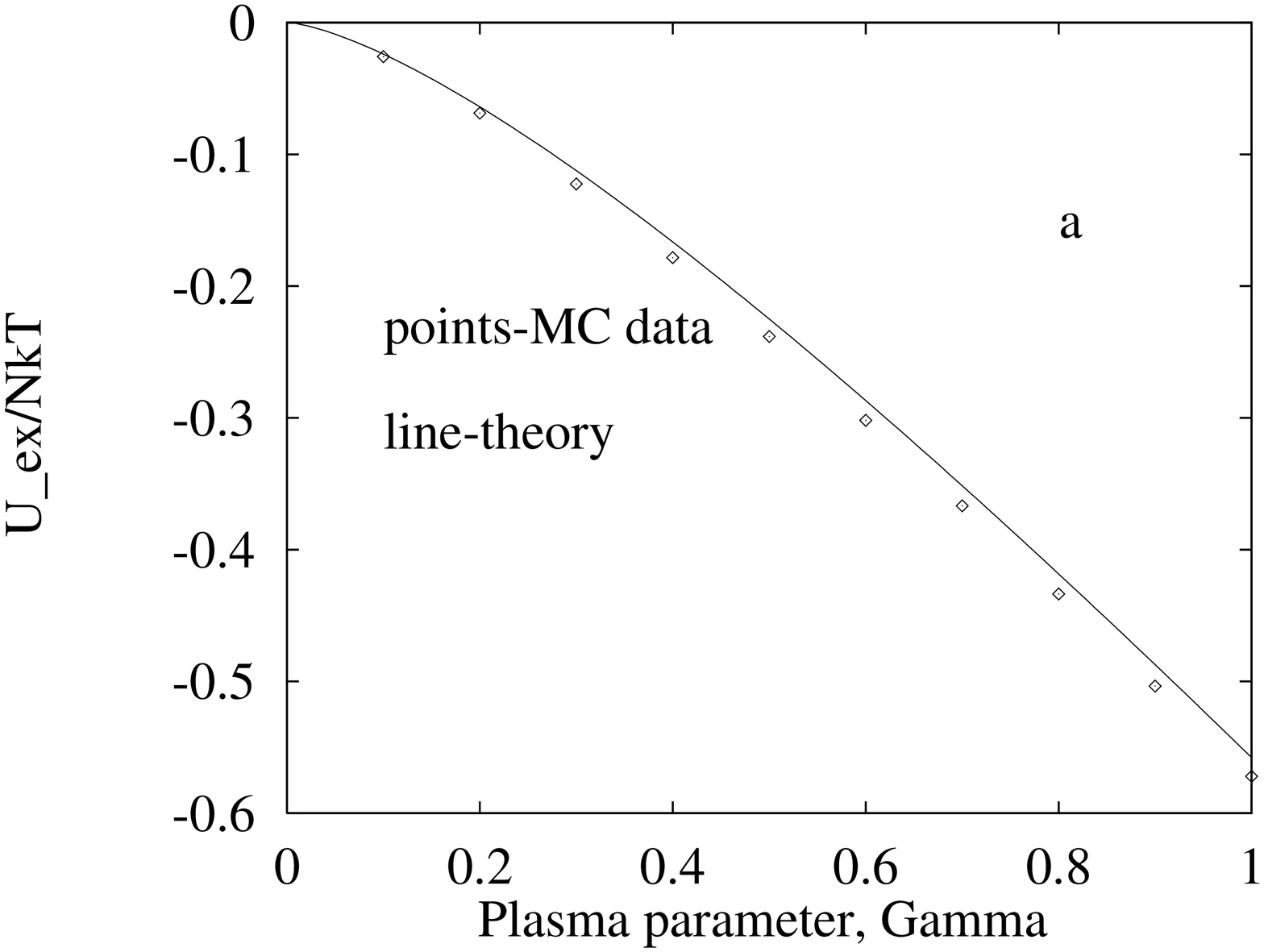,width=7.5cm,angle=270}
  \psfig{file=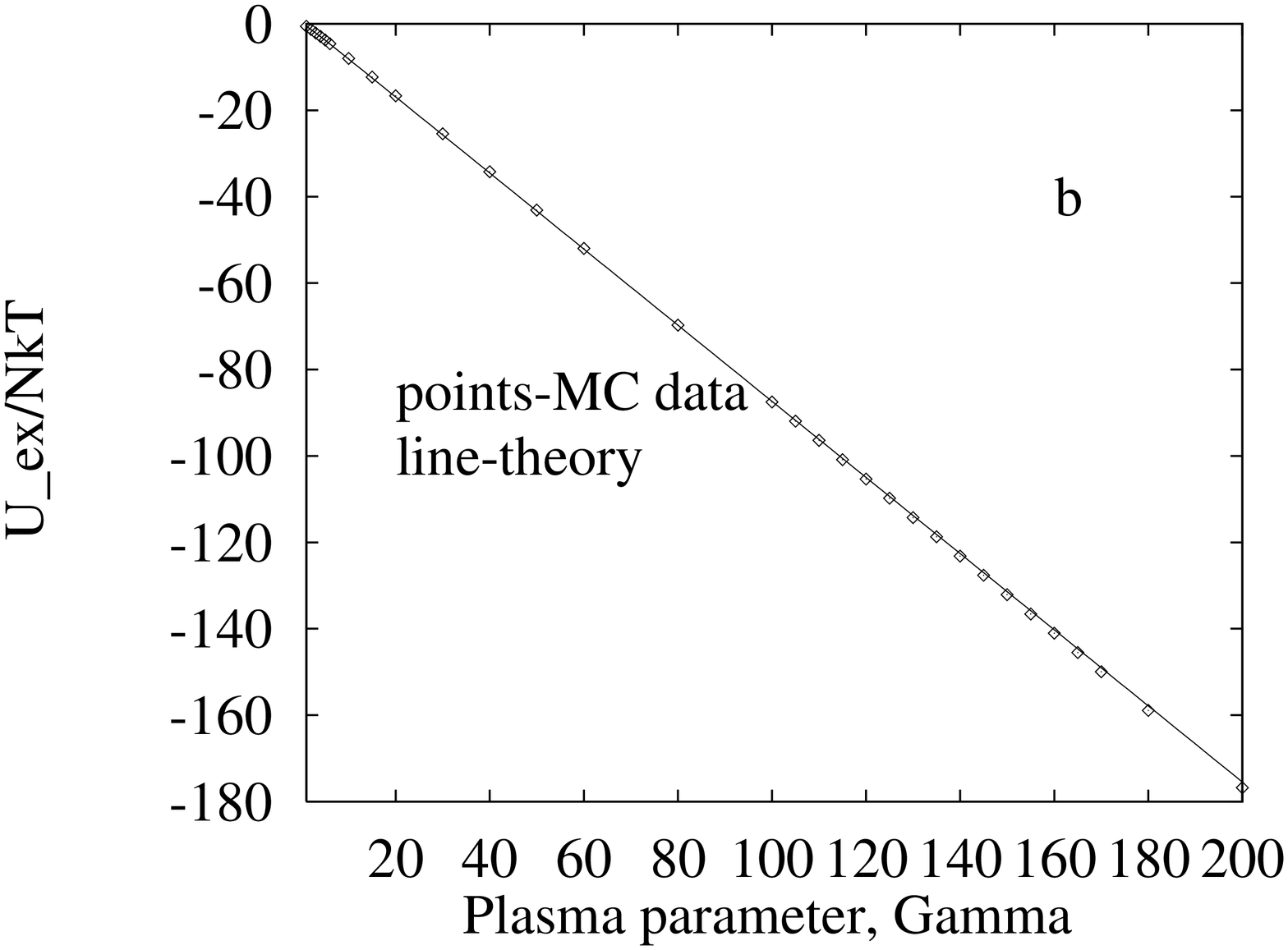,width=7.5cm,angle=270}}
%,height=4.8cm}}
\vspace{1.0cm}
\caption{ Shows the dependence of the excess internal energy 
of the OCP  
$U_{\rm ex}/Nk_BT$ on the plasma 
parameter $\Gamma=l_B/a_c$ 
($l_B=e^2/k_BT$, $a_c^{-3}=4\pi\rho/3$). 
Points give the Monte Carlo data (Ref.~[14,18] for 
$0.1 \le \Gamma \le 1$ and Ref.~[6,20] for $\Gamma>1$); 
lines represent the theoretical result,  Eq.(\ref{resU}) }
  \label{fig:1}
\end{figure}
%\end{minipage}

%\vspace{1.0cm}

%\vspace{1.0cm}

%\begin{minipage}{16cm}
\begin{figure}[htbp]
  \centerline{ 
  \psfig{file=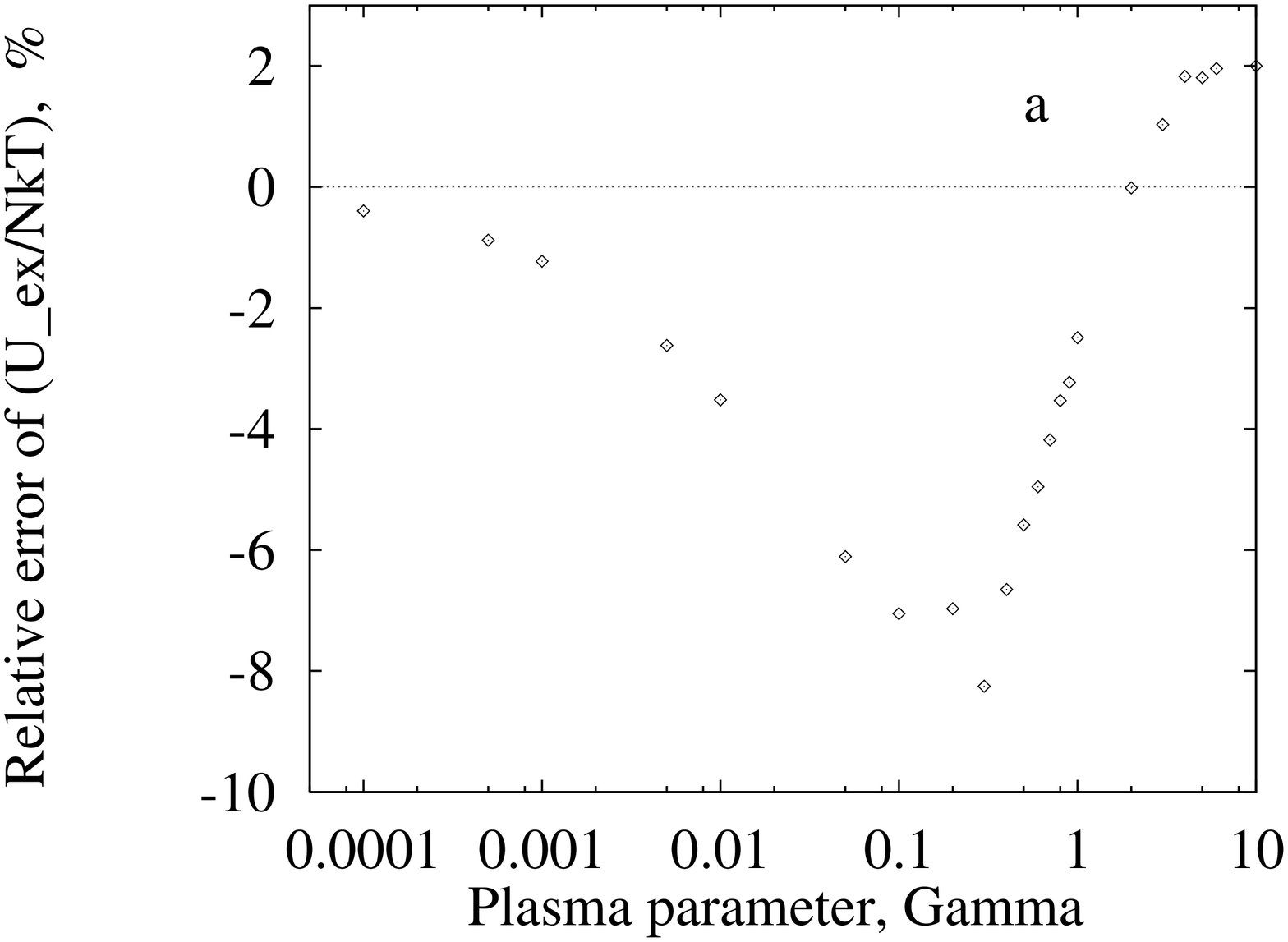,width=7.5cm,angle=270}
  \psfig{file=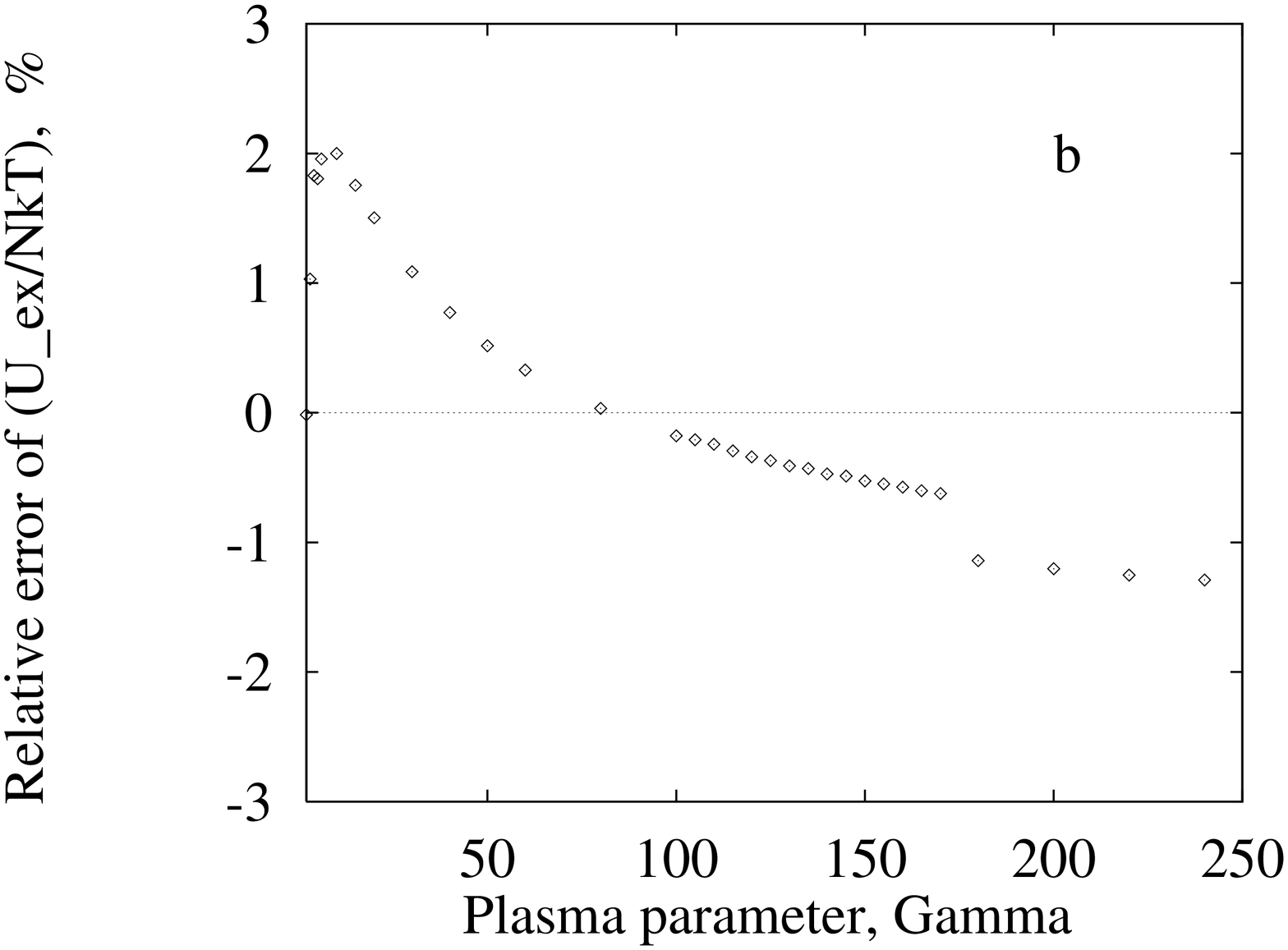,width=7.5cm,angle=270}}
%,height=4.8cm}}
\vspace{1.0cm}
\caption{ 
Shows the relative error (\%) of the analytical expression 
(\ref{resU}) for the excess internal energy 
of the OCP $U_{\rm ex}/Nk_BT$ as a function of the plasma parameter 
$\Gamma$. The Monte Carlo data  of Ref.~[14,18] for 
$0.1 \le \Gamma \le 1$ and of Ref.~[6,20] for $\Gamma >1$ are used. For 
$\Gamma <  0.1 $  Eq.(\ref{resU}) is compared with the Abe expansion 
Eq.(\ref{abeU})}. 
  \label{fig:2}
\end{figure}
\end{minipage}

%\vspace{1.0cm}
\noindent 
In particular the equation of state in a  form of the 
virial expansion may be recovered \cite{ort}. 
Unfortunately this expansion 
does not provide the closed analytical expression for the excess 
thermodynamic functions, which may be used with acceptable accuracy  
for all the range of $\Gamma$.

Thus, dealing with a problem where the relevant range of 
plasma parameter is not known in advance  one 
should preferably use the simple closed-form  equations suggested in the 
present study.

\section{ CONCLUSION}

A "first-principles" equation of state for the one-component plasma is 
derived that has a correct Debye-Huckel behavior at the limit of  
small plasma parameter $\Gamma$ and demonstrates a linear dependence on 
$\Gamma$ at $\Gamma \gg 1$. The obtained 
coefficient $0.881$ at the linear leading term 
is close to the corresponding coefficient $0.899$ found in the 
Monte Carlo simulations. The simple analytical expression for 
the excess internal 
energy reproduces the MC data  within  $1-2\%$ accuracy 
for the most of  range of $\Gamma$  ($0 \le \Gamma \le 250$) and has  
a typical deviation  of the order of $2-5\%$ 
(with a maximal one  $\approx 8 \%$) for $0.01 < \Gamma < 1$. 

\section{Acknowledgements}
Helpful discussions with John Valleau, George Stell, Werner Ebeling, 
Yaakov Rosenfeld and 
Jens Ortner are highly appreciated. The author is thankful for Torsten 
Kahlbaum for providing  some important references. Financial support 
of NSERC of Canada is acknowledged.


\begin{references}
\bibitem{remarkid}
         The term with ${\bf k}=0$ conceals with the terms 
         describing the self-interaction of the compensating 
         background and interaction of the background with particles.
\bibitem{remarkdiv} 
         The self-energy is equal to the energy of interactions of 
         two charges at vanishing distance. This   
         divergency does not affect the final result and may be 
         removed considering the lattice version of the 
         problem \cite{khol,klein}, or (equivalently) implying an 
         infinitesimally 
         small hard-core for interacting particles.
\bibitem{remark1}
         Note that the restriction $\varphi_{-{\bf k}}=\varphi^*_{{\bf k}}$
         does not reduce the total number of {\em physically  different} 
         modes, since the modes with the wave-vectors ${\bf k}$ and 
         $-{\bf k}$ correspond to that  spreading in the 
         opposite directions.
\end{references}
\end{document}